\documentclass[12pt,a4paper]{article}

\usepackage{graphicx}
\usepackage{times}
\usepackage{indentfirst}
\usepackage[tiny]{titlesec}

\newenvironment{keywords}
  {~\vspace{.6cm}\begin{flushright}
    \fontsize{10pt}{12pt}
  }
  {\end{flushright}}

\newenvironment{itcauthor}
  {\vspace{.8cm}\begin{flushleft}}
  {\end{flushleft}}

\newenvironment{itctitle}
  {
    \begin{center}
    \fontsize{13pt}{15pt}
    \vspace{1cm}
  }
  {
    \vspace{1cm}
    \end{center}
  }

\newenvironment{itcabstract}
  {
    \begin{list}{}{\setlength{\leftmargin}{1cm}}
    \item
    \fontsize{10pt}{12pt}
  }
  {
    \end{list}
  }

\footnotesize{
  \fontsize{10pt}{11pt}
}
\makeatletter
\renewcommand{\@makefntext}[1]{\noindent\@makefnmark#1}
\makeatother

\titleformat{\section}{\bfseries}{\bfseries\thesection.}{.4cm}{}
\titlespacing{\section}{0pt}{26pt}{14pt}

\titleformat{\subsection}{\normalfont}{\normalfont\thesubsection.}{.3cm}{}
\titlespacing{\subsection}{0pt}{14pt}{14pt}

\titleformat{\subsubsection}{\normalfont}{\normalfont\thesubsubsection.}{.3cm}{}
\titlespacing{\subsubsection}{0pt}{14pt}{0pt}

\newcounter{itccaption}
\usepackage{amssymb}
\usepackage{graphicx}
\usepackage{verbatim}
\usepackage[latin1]{inputenc}
\usepackage{paralist}
\usepackage{subfigure}
\usepackage{times}
\usepackage{sidecap}
\usepackage[footnotesize]{caption}
\begin{document}
\thispagestyle{empty}


\begin{itcauthor}
  Martin BURKHART\footnote{Computer Engineering and Networks Laboratory, ETH Zurich, Switzerland, burkhart@tik.ee.ethz.ch},
  Daniela BRAUCKHOFF\footnote{brauckhoff@tik.ee.ethz.ch},
  Martin MAY\footnote{may@tik.ee.ethz.ch},
\end{itcauthor}

\begin{itctitle}
On the Utility of Anonymized Flow Traces for Anomaly Detection
\end{itctitle}

\begin{itcabstract}
The sharing of network traces is an important prerequisite for the development
and evaluation of efficient anomaly detection mechanisms.
Unfortunately, privacy concerns and data protection laws prevent
network operators from sharing these data. Anonymization is a promising 
solution in this context; however, it is unclear if the sanitization of data preserves the 
traffic characteristics or introduces artifacts that may falsify traffic analysis results.
In this paper, we examine the utility of anonymized flow traces for anomaly detection.
We quantitatively evaluate the impact of IP address anonymization,
namely variations of permutation and truncation, 
on the detectability of large-scale anomalies. 
Specifically, we analyze three weeks of un-sampled and non-anonymized 
network traces from a medium-sized backbone network. 
We find that all anonymization techniques, except prefix-preserving permutation, degrade the utility 
of data for anomaly detection. We show that the degree of degradation 
depends to a large extent on the nature and mix of anomalies present in a trace.
Moreover, we present a case study that illustrates how traffic characteristics 
of individual hosts are distorted by anonymization.
\end{itcabstract}

\section{Introduction}

One of the principal reasons for the slow progress in anomaly detection research 
is the lack  of publicly available, unaltered network traffic traces.
The sharing of traffic data is hindered since releasing 
data always introduces a threat to users' privacy. 
Even when data
export is restricted to packet headers, as is the case with Cisco
NetFlow, a certain amount of personal information may still be
extracted and exploited to illegitimately profile user behavior. This
threat has already been recognized by data protection legislation in both
Europe \cite{eu-privacy1,eu-privacy2} and the United States~\cite{Ohm:2007:Legal}. 
As a result, multiple anonymization tools
that prevent the leakage of privacy information have been developed, such as 
FLAIM~\cite{Slagell:2006:Flaim}, TCPdpriv~\cite{tcpdpriv}, and CryptoPAn~\cite{Fan:2004:Prefix}.

For practical reasons these tools have been evaluated only with
regard to privacy concerns (e.g., in
~\cite{Coull:2007:Inferring,Koukis:2006:Risks,Brekne:2005:Attacks,Ribeiro:2008:Prefix}). The remaining question is whether the data sanitization 
preserves the traffic characteristics, or introduces artifacts that compromise the utility
for research purposes.
For researchers and engineers with access to unanonymized data sets, this is
not an issue; unfortunately, only few research institutes have such traffic traces available
and the large majority works with publicly available, but already anonymized data sets.
As a result, a study on the impact of anonymization methods is essential and indeed overdue, 
since numerous anomaly detection algorithms have been
evaluated with anonymized data. Therefore, the goal of this work
is to determine to what extent the anonymized traces falsify the
results of commonly used anomaly detection mechanisms.

To the best of our knowledge, the specific problem we are investigating 
has not yet been addressed in literature. 
In \cite{Soule:2007:Anomalies}, Soule et al. study
NetFlow data from two backbone networks that apply different
sampling and anonymization schemes, and suggest that anonymization
might have an impact on anomaly detection. The problem of data loss due to 
anonymization is also identified in~\cite{Zhang:2006:Outsourcing}, where the authors give qualitative recommendations for
anonymization of NetFlow logs when security services are outsourced.
Yurcik et al.~\cite{yurcik2008pat} analyze single-field
anonymization tradeoffs with regard to intrusion detection.
Unfortunately, their dataset contains already anonymized IP
addresses, hence the impact of IP address anonymization techniques
on utility is not studied. On the contrary, \cite{Brauckhoff2005}
and \cite{Mai2006} studied the utility of sampled traffic traces
without addressing the effect of anonymization.

This paper represents the first comprehensive study on the utility
of anonymized data for statistical anomaly detection approaches. 
Specifically, we focus on the anonymization of IP addresses as they 
comprise the biggest threat to user privacy. We analyze the most 
popular IP address anonymization techniques, namely blackmarking,
truncation, random permutation, and (partial) prefix-preserving permutation.
Our contributions are as follows: 
(i) we introduce a generic methodology for evaluating the impact of 
anonymization on flow-based traffic analysis applications;
(ii) we quantify the utility of anonymized data for backbone anomaly
detection with the help of a three-week long data set from a medium-size
ISP and an anomaly detector based on a Kalman filter; and (iii) we 
present an overall estimate for the impact of anonymization
on the analysis of individual hosts' traffic characteristics.

To tackle the problem of determining the utility of anonymized data, we first 
introduce the granularity design space for traffic analysis. This design space
has two dimensions: the \emph{subset size} of the address space under investigation and 
the \emph{resolution} of the examination. 
We argue that the complete granularity design space is valuable for traffic analysis,
even though today, only a subset of it is used (e.g., Origin-Destination flows).
We then show how the granularity design space is diminished by 
the considered anonymization techniques.

In section \ref{sec:methodology} and \ref{sec:results}, we present a measurement study on the impact 
of anonymization on the data utility for anomaly detection. We introduce the Kalman filter used for anomaly detection
and evaluate the detectors' performance on the unanonymized and unsampled, manually labeled three-week data set.
We assess the utility of anonymized data by evaluating the detector performance on the
restricted set of metrics available with each anonymization scheme. Specifically, we evaluate
the results with the help of ROC curves~\cite{fawcett2006ira} that plot false positives vs. true positives for a 
range of thresholds.
We show that the restriction of available granularities
through anonymization degrades the performance of anomaly detection.
Surprisingly, we find that the degree of degradation differs for volume, scan/DoS, and network
fluctuation anomalies, as well as for UDP and TCP traffic. As an example, we will show that
for the detection of network fluctuations, the utility drops from 87\% to 70\% when
the number of truncated bits is doubled.

In section \ref{sec:example}, we study the effect of anonymization on detailed traffic analysis and root cause identification. 
We are able to show that even a small restriction of the subset size heavily impacts the visibility of anomalies. 
Finally we discuss our findings and conclude the paper in section \ref{sec:conclusion}.

\section{Utility of Anonymized Data for Anomaly Detection}
\label{sec:utility}

In this section, we investigate how five popular IP address
anonymization techniques impact statistical anomaly detection on flow data.
To make our discussion more systematic, we introduce a granularity design space for traffic analysis, 
and show how the different anonymization techniques diminish this
design space. 

\begin{figure*}[t]
  \centering
  \begin{minipage}[b]{0.45\textwidth}
    \centering
	  \includegraphics[scale=0.43]{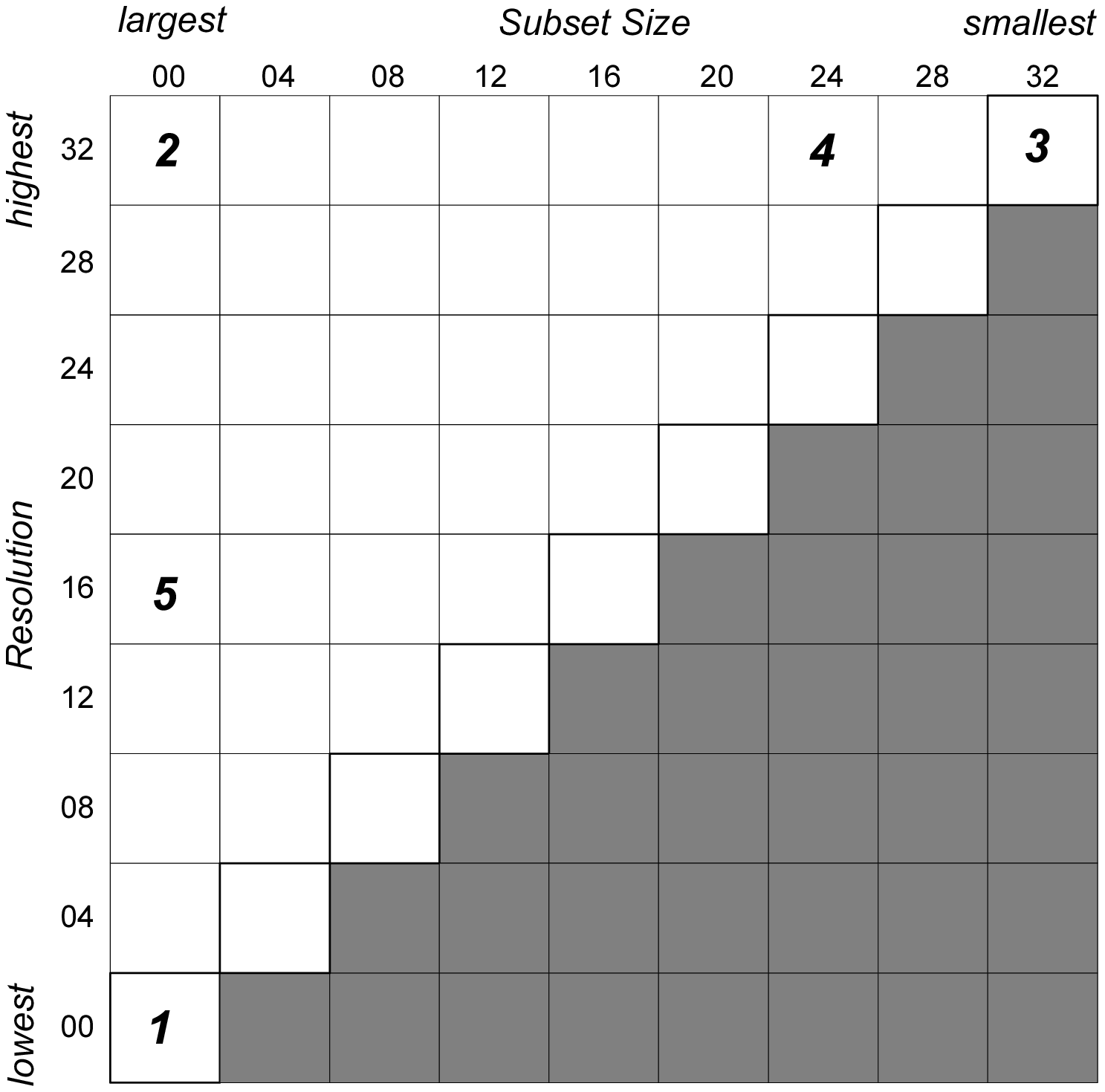}
    \caption{Granularity design space for metrics used in statistical anomaly detection.}
	  \label{fig:dspace}
  \end{minipage}
  \hspace{0.5cm}
  \begin{minipage}[b]{0.47\textwidth}
	  \centering
	  \includegraphics[scale=0.4]{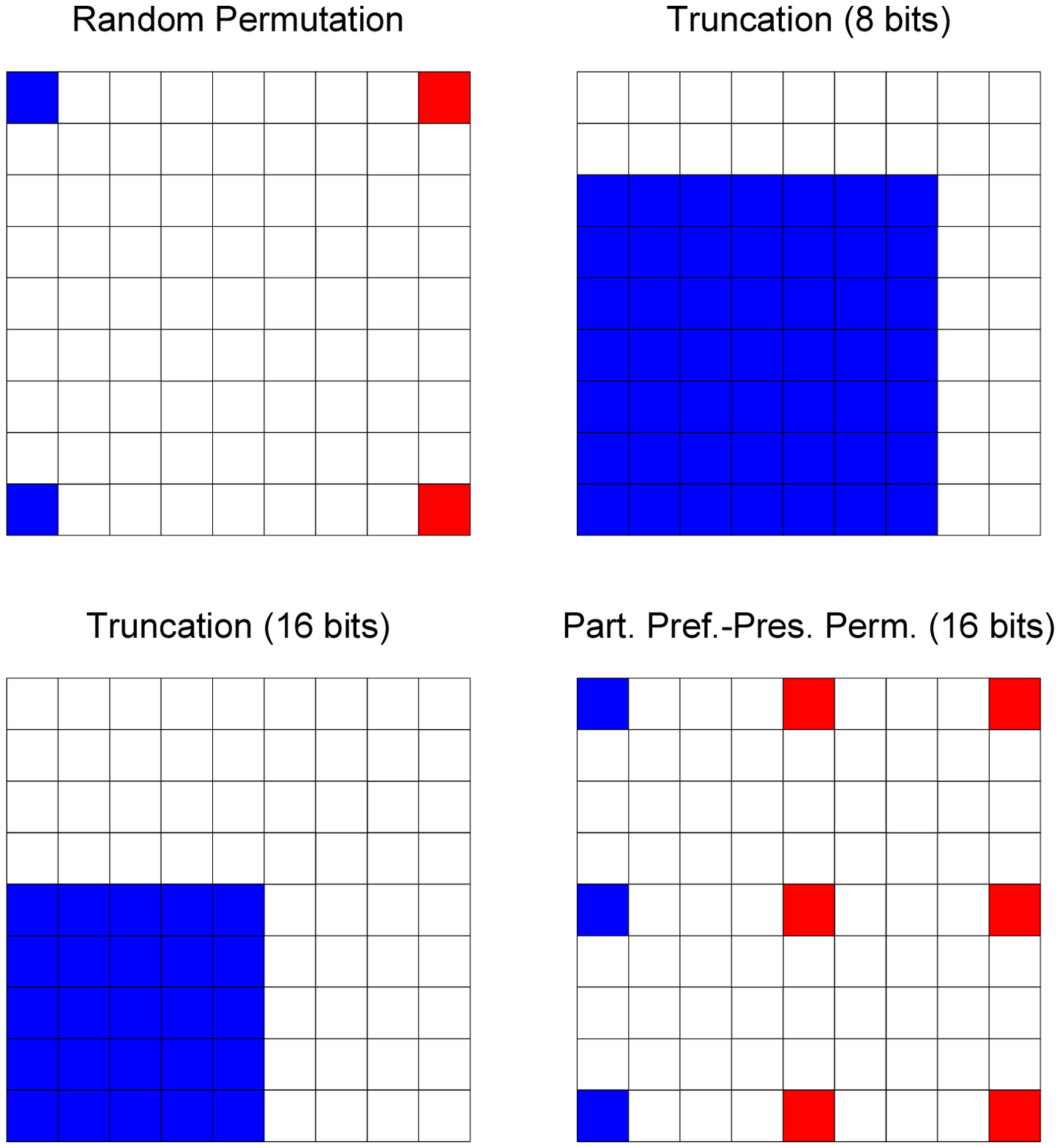}
	  \caption{Resolutions and subset sizes available with different anonymization techniques.}
	  \label{fig:manon}
	 \end{minipage}
\end{figure*}

\subsection{A Granularity Design Space}

Up to now the metrics used for anomaly detection, and traffic analysis in general, have been designed in an ad-hoc manner,
based on (i) the characteristics of the data set under study, and (ii) the 
type of traffic characteristic one is interested in. Prominent examples of such 
metrics are the well-known volume metrics, such as byte, packet, and 
flow counts, which are simply computed over all traffic in a given trace. Lakhina et al. in 
\cite{SubspaceMethod04} group the \emph{anonymized} traffic from the Abilene network into 
Origin-Destination (OD) flows before analyzing it further with Principal Component Analysis. Whereas, in \cite{Wei2006, Xu2005} 
host-based metrics such as IP address entropy or the number of active connections are used 
for host profiling in a clustering algorithm.

To investigate the impact of anonymization in a systematic manner, however, 
it is necessary to explore the whole granularity design space for traffic analysis.
To motivate the design space it is helpful to use an allegory from image processing or 
photography. When taking a picture, one focuses on the object of interest and 
selects a fine or coarse resolution to display the details at the desired level.
Similarly, the granularity design space has two dimensions: 
\begin{description}
\item[Subset size] The size of the 
network that is to be analyzed. When analyzing backbone data for example, we can 
analyze all traffic, or we can focus on a specific subnet in the backbone. 
\item[Resolution] The address granularity at which the traffic is analyzed.
We can select a very high resolution of IP addresses if one is interested
in profiling hosts, or a low resolution of Autonomous Systems 
(or OD flows) if one is interested in more global events.
\end{description}

We give a matrix representation of the granularity design space in Fig. \ref{fig:dspace}.
The x-axis ranges from the largest subset size of all traffic (00 bits),
to the smallest possible subset size of a single IP address (32 bits).
Likewise the y-axis ranges from the highest resolution of individual IP addresses (32 bits) to 
the lowest resolution of all traffic (00 bits).
The design space matrix can be divided in three sections: the upper triangle, 
the diagonal, and the lower triangle. Traditional volume metrics fall in one of the cells on the diagonal, 
where the subset size equals the resolution
(e.g., we select all traffic from a /16 subnet and compute the flow
counts over all traffic coming from that /16 network). The upper triangle features metrics where the 
resolution level is smaller than the size of the selected traffic subset. Such metrics, e.g, the number of 
unique IP addresses in the inbound traffic to a selected /24 subnet, are also frequently
used for traffic analysis. Finally, for metrics on the lower triangle, the resolution level is larger 
than the size of the selected traffic subset. Such metrics are rarely used today, and are 
thus of less interest to our study. 

Note, the full design space is not always available. For example, when working with data from stub networks, 
e.g., a campus network, where the maximum available subset size equals the IP address range
assigned to the studied network. For a /16 campus network the subset size may range from 16 to 32, 
i.e., subset sizes larger than 16 bit are not available. 
Nevertheless, to keep the subsequent discussion general we assume the whole design space is available.

To further illustrate the granularity design space, we give example metrics for five selected 
cells of the matrix:

\begin{itemize}
\item Cell 1 [00,00]: Select all traffic and set the resolution to 
the minimum. An example metric is the well-known volume over all traffic.
\item Cell 2 [00,32]: Select all traffic and set the resolution to
the maximum. Examples are the volume per IP address or the number of unique IP addresses
in all traffic.
\item Cell 3 [32,32]: Select traffic to/from one IP address and set the resolution 
to the maximum. Metrics falling in this category are, e.g., the number of unique ports
per IP address, or the number of unique IP addresses that the host under observation sends
traffic to.
\item Cell 4  [24,32]: Select traffic to/from one /24 network and set the resolution 
to the maximum. Examples for this case are the flow count per IP address, or the unique 
number of IP addresses that send traffic in the monitored /24 network.
\item Cell 5 [00,16]: Select all traffic and set the resolution to /16 networks.
An example metric is the volume per /16 networks in all traffic.
\end{itemize}

\subsection{How Anonymization Diminishes the Design Space}
\label{sec:ad}

\begin{table}[t]
    \centering
    	\scalebox{0.9}{
        \begin{tabular}{|c|c|c|c|c|c|}
        \hline
           IP  Address &  Truncation   &   Random  &  Prefix-Pres. & Partial Prefix-Pres. \\
            										 & (16 bits)   & Permutation   & Permutation  & Permutation (16 bits) \\
        \hline
           129.132.91.35   & 129.132.0.0   & 112.4.23.73    & 22.5.99.76   &  73.9.8.1   \\
        \hline
           129.132.91.177  & 129.132.0.0   &  62.12.96.67   & 22.5.99.41  &   73.9.181.17\\
        \hline
           129.132.8.37   & 129.132.0.0   & 205.72.5.18       & 22.5.181.92 &  73.9.1.230 \\
        \hline
           152.88.3.90       & 152.88.0.0    & 2.14.12.133        & 110.27.20.1 &  18.7.18.133  \\ 
        \hline
           152.96.99.2      &  152.96.0.0   & 19.0.111.20   & 110.9.0.12  & 24.125.43.6 \\
        \hline
           82.130.102.115      & 82.130.0.0   & 12.171.92.3       & 145.21.5.19  &  145.213.2.77 \\
        \hline
        \end{tabular}
      }
    \caption{Examples of IP address anonymization}
    \label{tab:Examples}
\end{table}

In the following we outline the studied IP address anonymization techniques
and show how they diminish the granularity design space.
The most commonly employed IP address anonymization techniques are
blackmarking, truncation, random permutation, prefix-preserving permutation, and partial prefix-preserving permutation.
An illustrative example for each technique, except blackmarking which is trivial, is given in Table~\ref{tab:Examples}.

The available subset of the design space for different anonymization techniques
is illustrated in Fig.~\ref{fig:manon}, where filled squares mark the possible combinations
of subset size and resolution for 
each anonymization technique. Note that for permutation-based approaches all fields with a subset
size smaller than 00 are marked with a different color. We did this to signify that
subsets of smaller sizes may be distinguished, but not identified, since the mapping from real to 
anonymized IP addresses is usually not known. Hence, a subset of interest has to be identified 
by different means, e.g., by selecting subnets with particular traffic characteristics. For the subsequent
analysis, however, we make no distinction between the two cases.

\emph{Blackmarking (BM)} is the simplest of all studied anonymization techniques. 
It blindly replaces all IP addresses in a trace with the same value. As a result, 
all information about individual IP addresses or subnets is lost and only metrics with the 
lowest resolution and the largest subset size, e.g., the volume over all traffic, can be computed. 
This corresponds to a single cell in the design space matrix, the lower left corner of the matrix.
Several traces from the Internet traffic archive (LBNL) are anonymized with blackmarking.
Please refer to the UCRchive for a comprehensive list of available traces
\footnote{http://networks.cs.ucr.edu/ucrchive/measurement.htm}.

\emph{Truncation (TR\{t\})} replaces the $t$ least significant bits
of an IP address with 0. Thus, truncating 8 bits would replace an
IP address with its corresponding class C network address. 
With respect to the design space, this means only metrics with a resolution and 
subset size of [00, 32 - t] can be computed when truncation is used. 
The number of available granularities decreases
with $t$, the number of truncated bits, as illustrated in Fig.~\ref{fig:manon}.
The traces from the Abilene network, which have been used to evaluate numerous anomaly
detection approaches, are anonymized with truncation of 11 bits.

\emph{Random permutation (RP)} translates IP addresses using a
random permutation that does not preserve the prefix structure. Since
permutation creates a one-to-one mapping, the number of distinct IP
addresses is the same. Hence, when random permutation is used for anonymizing a trace, 
metrics that can be computed on it may only feature the highest 
and lowest resolution values, as well as largest and smallest subset sizes 
(see Fig.~\ref{fig:manon}). Note that these correspond to the four corners of the design space matrix.
A special case of random permutation is the renumbering of IP addresses (e.g., TCPdpriv with level 0).
Packet Traces from UCLA CSD, as well as several traces from the Internet traffic archive (LBNL) 
are sanitized using random permutation.

\emph{Partial prefix-preserving permutation (PPP\{p\})}, as proposed in~\cite{Pang:2006:tcpmkpub},
permutes the host and network part of IP addresses independently. It preserves the prefix structure in a trace 
at one specific prefix length $p$, and at the level of IP addresses. 
Consequently, this technique retains all granularities that have a resolution and subset size of 
either 00, $p$, or 32 (see Fig.~\ref{fig:manon}). PPP is a popular technique that is used for anonymizing traces 
from the Passive Measurement and Analysis project (PMA) and the Internet traffic archive (LBNL).
PMA uses PPP\{12\} and PPP\{16\}. Moreover, level 1 of TCPdpriv corresponds to PPP\{16\}.

\emph{Prefix-preserving permutation (PP)} permutes IP addresses so that two addresses sharing a common
real prefix, also share an anonymized prefix of equal length (see e.g.,~\cite{Fan:2004:Prefix}). 
This is actually the best anonymization technique with respect to utility since it preserves the 
full design space. We will use it in our measurement study as a reference to a perfect anonymization scheme (with respect to utility). 
PP is applied to traces from CAIDA and CRAWDAD.

Note that anonymization always involves a tradeoff between data utility and the risk of privacy violations~\cite{Duncan:2001:RU-Map}. Ideally, an anonymization scheme would guarantee perfect protection from privacy violations (low risk) without affecting the utility of data with respect to some target 
application (high utility). In this paper, however, we focus on data utility only. For attacks on anonymization techniques please refer to~\cite{Coull:2007:Inferring,Koukis:2006:Risks,Brekne:2005:Attacks,Ribeiro:2008:Prefix}.

\section{Methodology}
\label{sec:methodology}

In this section, we describe our methodology for studying the
impact of anonymization on statistical anomaly detection.
We introduce the data set used in this study, and
describe the methodology for classifying it. We further 
present the Kalman filter that is used as detection algorithm.

\subsection{Measurement Data}\label{sec:DataSet}


The data used in this study was captured from the four border routers of the
Swiss Academic and Research Network (SWITCH, AS 559)~\cite{Switch},
a medium-sized backbone operator, connecting several universities
and research labs (e.g., IBM, CERN) to the Internet. The SWITCH IP
address range contains about 2.4 million IP addresses and the
traffic volume varies between 60 and 140 million NetFlow records per
hour. We analyzed a three-week period (from August 19th to September 10th
2007). This data set contains a variety of anomalies with
diverse characteristics. In total, 43.2 billion flows covering a
volume of 713 Terabytes of traffic were analyzed.
In contrast to previous work, this study is based on un-sampled and
non-anonymized flow data. Such datasets are difficult to obtain (at least 
over longer observation periods), but mandatory if bias and distortion
in the results are to be avoided.

\subsection{Ground Truth}

The first step of any measurement study on anomaly detection is
the establishment of ground truth for the available traces.
Unfortunately, obtaining ground truth for an unclassified data set is still a large
challenge and involves a lot of manual inspection. 
In the following, we describe our methodology for labeling the dataset.

\emph{Visual inspection of metric timeseries}: We computed the timeseries for five well-known 
metrics over an entire three weeks period at 15-minute intervals, resulting in 2016 data points per metric. 
As metrics, we selected byte, packet, and flow counts, unique IP address counts, and the Shannon 
entropy\footnote{$H(X) = - \sum^n_{i=1}{P(x_{i}) \cdot log_{2}\bigl( P(x_{i}) \bigr)}$} of flows per IP address.
Moreover, we distinguished incoming and outgoing traffic, as well as TCP and UDP traffic adopting what is common practice
in the anomaly detection community. Finally, we visually inspected all these timeseries for unusual events.

\emph{Analysis of raw NetFlow traces}: For all intervals that could not be classified 
by timeseries inspection with high confidence, we did further analysis on the raw NetFlow traces.
For this purpose, we used nfdump~\cite{NFDUMP}, a tool developed by
SWITCH for forensic analysis to collect more information about suspicious events, 
e.g., which hosts and ports are affected.

\emph{Assigning ground truth to each interval}: If at least \emph{one} of the analyzed metric 
timeseries exposed an unusual event in some interval, we classified that interval
as anomalous. Note here that most events were visible across multiple metric timeseries.


\begin{SCtable}
    \centering
    \scalebox{0.9}{
    \begin{tabular}{|l||l|l|l|l|l|l|}
            \hline
        & Vol & DoS & Sca & Flu & Unk & Tot \\
        \hline
        \hline
        TCP & 75 & 32 & 539 & 24 & 19 & 689\\
        \hline
        UDP & 64 & 14 & 4 & 239 & 28 & 339\\
        \hline
    \end{tabular}
    }
    \hspace{0.5cm}
    \caption{Ground truth: Number of anomalous intervals per anomaly type and total for UDP/TCP.}
    \label{tab:anomalies}
\end{SCtable}

\emph{Identifying the anomaly type}: Having classified all intervals as normal or anomalous, 
we went one step further and assigned the anomalous events
to different types. Since a commonly agreed methodology for
classifying known anomalies is not yet established, we define and distinguish 
the following types of events:
\begin{itemize}
\item {\bf Volume}: Volume anomalies are events that cause a sharp increase or decrease in the volume-based
metrics, but do not affect the feature-based metrics. In our trace, we found two large loss events and several high-volume flows
or alpha flows.
\item {\bf(D)DoS}: Denial of Service attacks cause a concentration of the flows on one or few target IP addresses and hence a
drop in the destination IP address entropy. If, on top, the attack is distributed, we will additionally see a spike in the
source IP address counts and entropy metrics. If the attack is large in terms of flows or even packets, in addition, it will
cause a spike in volume-based metrics.
\item {\bf Scan}: Scans provoke an increase in the destination IP address counts and entropy. If the attack
sources are distributed, we will also see an increase in the source IP counts.
\item {\bf Network Fluctuation}: Events that cause an increase or decrease in the IP counts at lower resolutions but are not
significant in the IP address counts at the highest resolution, fall into this class. 
Examples of such anomalies are ingress shifts and route flaps, but also massively distributed coordinated events 
that involve only a small number of IP addresses (e.g., botnet activity or stealth scans).
\item {\bf Unknown}: Despite the classification effort that was made, some events remained unclassified. All unclassified
events fall into this class.
\end{itemize}

Table \ref{tab:anomalies} summarizes the identified events in our three-week long trace for UDP
and TCP traffic. Note that we counted the number of anomalous intervals not the number of anomalies. Therewith
the number of anomalous intervals can be quite large for
anomalies that persist over several hours or even days. For example, a large part of the 542 TCP-scanning intervals belongs to
a single long-lasting event. Likewise, most of the 239 intervals classified as \emph{network fluctuation} belong to one single 
event that reappeared every 2 hours over several days, but lasted each time only for few intervals.

\subsection{Anomaly Detection with the Kalman Filter}

\begin{figure*}[t]
\centering

\begin{minipage}{0.47\textwidth}
	\centering
	\vspace{-0.2cm}
  \includegraphics[scale=0.61]{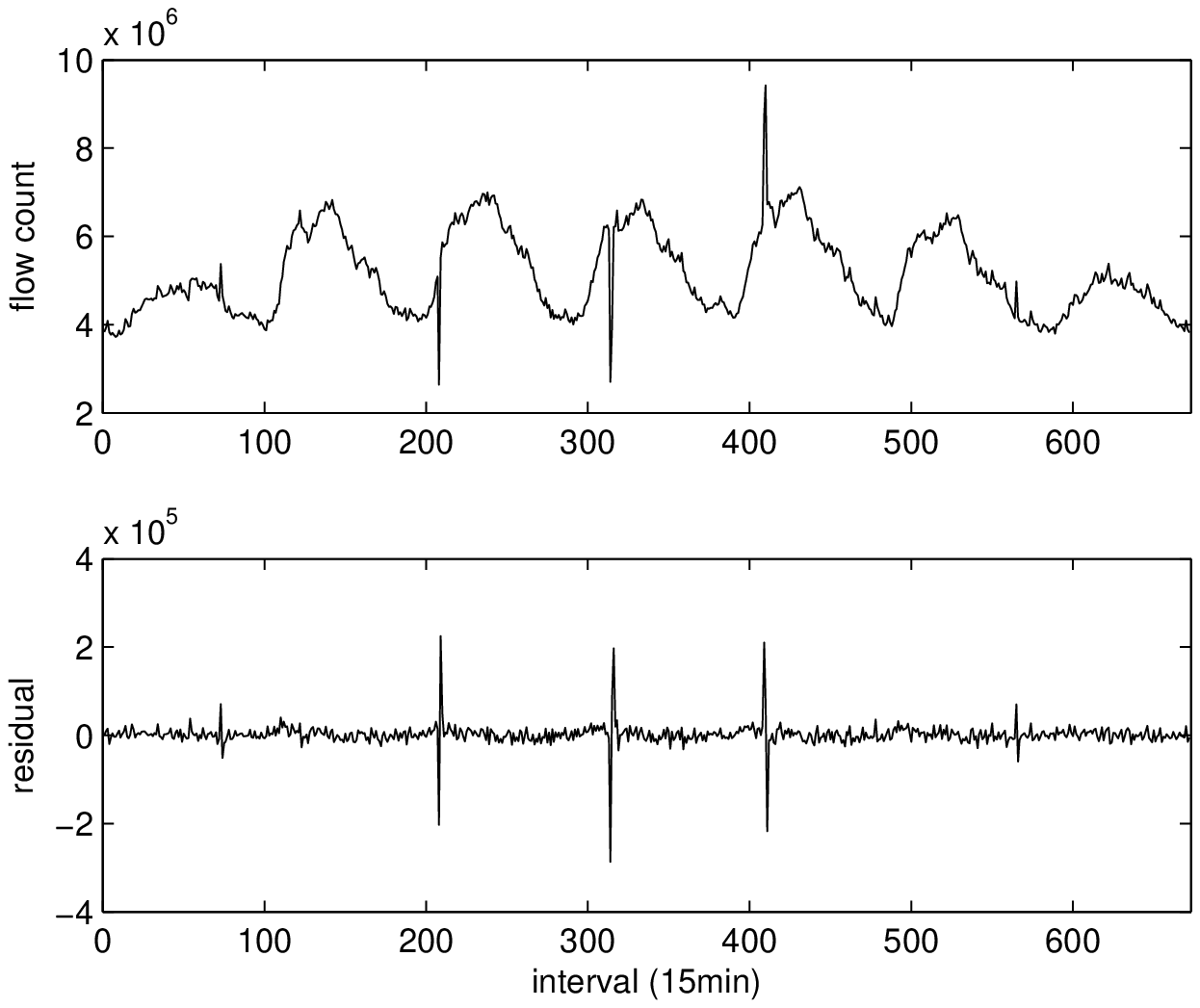}
  \caption{Time series and corresponding residual signal from the Kalman filter.}
  \label{fig:kalman}
\end{minipage}
\hspace{0.3cm}
\begin{minipage}{0.47\textwidth}
  \centering
  \includegraphics[angle=270, scale=0.47]{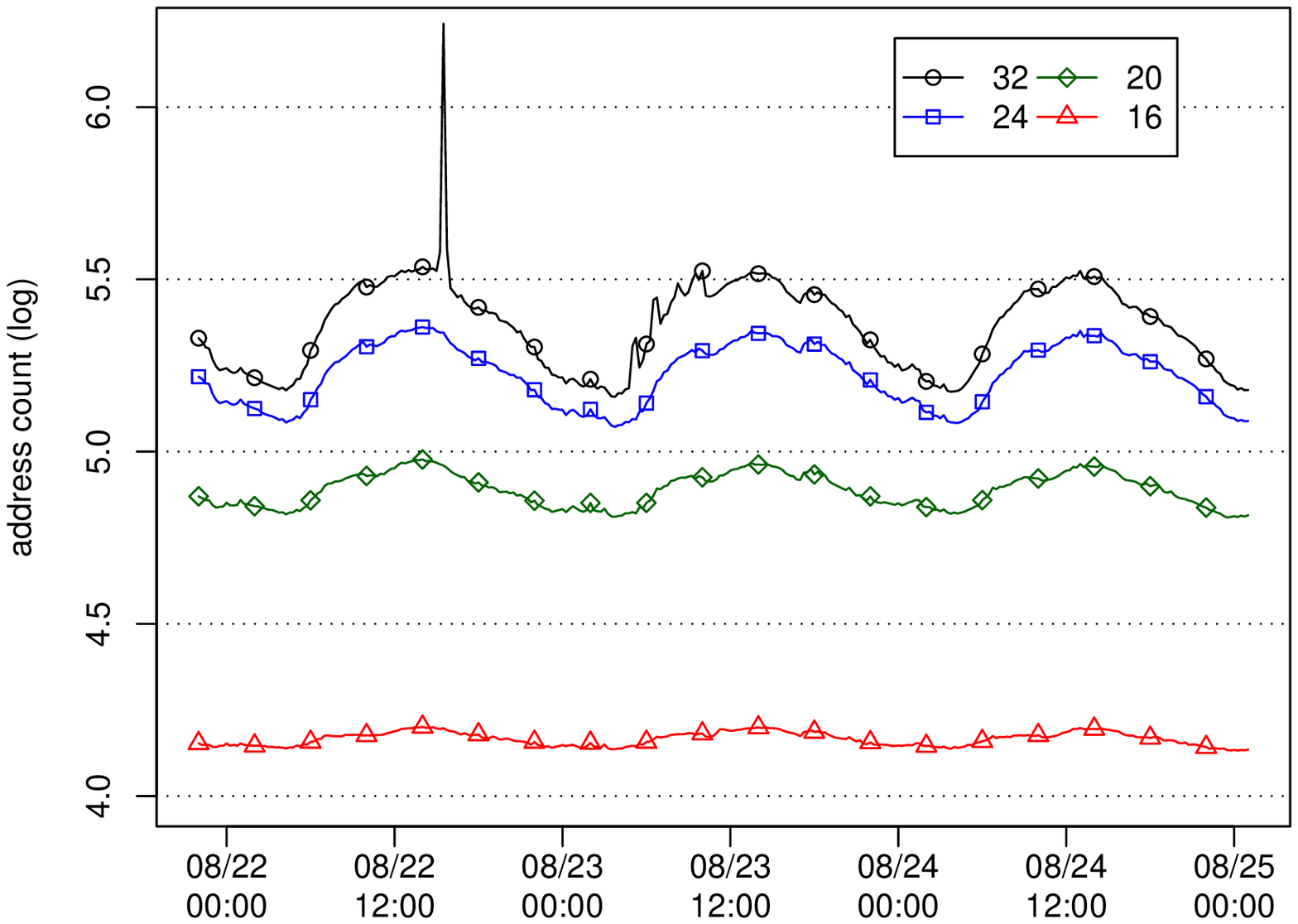}
  \caption{Illustration of the loss of resolution effect. The unique address count at resolutions 32, 24, 20 and 16 is shown.}
  \label{fig:lossOfResolution}
\end{minipage}

\end{figure*}

From the list of available statistical anomaly detection methods, we 
selected the Kalman filter since its excellent performance for anomaly detection has 
been shown in \cite{Soule2005}.
The Kalman filter is an efficient recursive filter that estimates the state of a dynamic
system from a series of incomplete and noisy measurements. It models normal traffic as a ``measurement-corrected'' AR(1) process plus zero-mean Gaussian noise. The difference between this model and the actual measured time series, the so-called residual, is used for detection (see Fig. \ref{fig:kalman} for an illustration). An alarm is raised by the detector if the residual excesses some threshold. 

We calculated a total of 60 metrics (see next Section) on the three-weeks data set, and applied the Kalman 
filter \emph{separately} to each of those. This results in a \{60 x 2016\} matrix of residual 
values, one for each metric and interval. An anomaly is detected if at least one of the 
60 residual values for an interval exceeds a threshold. 
We assess the performance of the Kalman filter with the help of ROC curves 
~\cite{fawcett2006ira}. ROC curves plot the rate of false positives against the rate of true 
positives for a range of thresholds. As thresholds, we use multiples of the a posteriori estimation for the standard deviation ($s$) of the considered metric. 
The thresholds range from $0.2s$ (top right corner) to $2.4s$ (bottom left corner).
Remind from theory that an interval which exceeds the standard deviation of the noise process 
is considered unusual. In general, the higher the true positive rate at a particular false 
positive rate, the better the performance of the detector. Hence, the curve of an optimal detector goes through the top left corner whereas
a curve close to the diagonal represents random guessing.

For both UDP (left) and TCP (right) traffic, Fig.~\ref{fig:rocall} shows one ROC curve per 
anomaly class, and one curve for the overall detection capabilities. We restrict our analysis 
to a specific type of anomaly as follows: we exclude all
intervals that have been manually classified as anomalous, but of a different type,
and used these shortened timeseries as input for the detector.
We see from this Figure that the Kalman filter generally works well. 
For UDP, we obtain very high true positive rates at a small false negative rate for all 
classes of anomalies. Detection results for TCP traffic are slightly worse. This is 
however expected since TCP has a larger traffic share than UDP and is also more volatile 
compared to UDP traffic.

When examining the traces in detail, we found that false positives are 
often due to fast increases or decreases in the normal daily traffic cycle, which are misinterpreted by the Kalman filter 
as anomalous events. Another source for false positives are temporary increases in the volatility of a metric.
This type of false positives can be avoided by recalibrating the Kalman filter each time the volatility changes.
We also observed that the Kalman filter tends to miss anomalies that increase or decrease more slowly, and take multiple intervals to grow 
to their full strength. For some of those, however, the Kalman filter detects the end of the anomaly when the traffic suddenly
falls or rises to its previous level.
There is one more thing to point out in Fig.~\ref{fig:rocall}: The false positive rate at a specific threshold 
is practically the same for all classes of anomalies (i.e., all markers representing the same threshold 
are more or less vertically aligned). The reason therefore is that the false positive rate depends only on the 
normal traffic, which is the same for all cases, and not on the studied anomaly. 

\subsection{Computing the Utility of Anonymized Data}

\begin{SCtable}
    \centering
    \scalebox{0.9}{
    \begin{tabular}{|c||c|c|c|c|}
            \hline
        & \small{vbm\{00\}} & \small{fbm\{16\}} & \small{fbm\{24\}} & \small{fbm\{32\}} \\
        \hline
        \hline
        \small{PP} & x & x & x & x \\
        \hline
        \small{PPP(16)} & x & x & & x\\
        \hline
        \small{RP} & x &  &  & x \\
        \hline
        \small{TR(08)} & x & x & x & \\
        \hline
        \small{TR(16)} & x & x & & \\
        \hline
        \small{TR(32)} & x & & & \\
        \hline
    \end{tabular}
    }
	  \hspace{0.5cm}
    \caption{Metrics available with different anonymization techniques (vbm = volume-based metrics,
    fbm = feature-based metrics).}
    \label{tab:metrics}
\end{SCtable}

Basically, we use the same methodology for the non-anonymized case as for the anonymized traffic.
The difference is, however, that we run the Kalman filter only on the subset of the 60 metrics that is 
available for the anonymization technique under study. 

The 60 studied metrics are different variants of three volume-based metrics (vbm) (byte, packet, and flow counts)
and two feature-based metrics (fbm) (the unique IP address count, and the 
Shannon entropy of flows per IP address). We distinguished TCP and UDP traffic as well as incoming 
and outgoing traffic. Moreover, we used a subset size of 0 for all metrics, i.e., we computed our metrics
over all available traffic. Since we were interested in exploring how the
restriction of available resolutions affects anomaly detection, we computed the
metrics at four representative resolution levels of \{00, 16, 24, 32\} bits\footnote{A resolution of 8 bits 
is too low for our data set.}. Here, we made a distinction
between volume- and feature-based metrics. We computed volume-based metrics only at the
lowest resolution of all traffic. This is because the computation of
volume metrics at higher resolutions (e.g., the volume per IP) results in one time series
per entity, and a clustering mechanism would be required to summarize them into one metric.
The impact of anonymization on clustering algorithms, however, is not subject of this study.
Feature-based metrics, on the other hand, were computed at a resolution of \{16, 24, 32\} bits. The lowest
resolution was not used for feature-based metrics since it results always in a value of one
(e.g., there is only one unique /0 prefix in the trace).
Therewith, we obtain a total of
$(3 [vbm] + (2 [fbm] \times 2 [src/dst] \times 3 [res])) \times 2 [in/out] \times 2 [udp/tcp] = 60 $
detection metrics.

The resolutions available with each anonymization scheme are 
given in Table \ref{tab:metrics}. Also refer to Fig.~\ref{fig:manon} which illustrates the 
subset of the design space available with each technique. The volume-based metrics computed at a 
resolution of 00 bits (vbm\{00\}) are available for all anonymization techniques.
Feature-based metrics computed at a resolution of 16 bits are available with
all techniques that retain this resolution, i.e., PP, PPP(16), TR(08), and TR(16).
Likewise, feature-based metrics at a resolution of 24 bits are available with 
all anonymization techniques that retain the resolution of 24 bits, i.e., 
PP, and TR(08). Finally, feature-based metrics computed at a resolution of 32 bits
are available with all permutation-based techniques since these retain the notion of 
individual IP addresses.

To assess the utility, we compared the ROC curves obtained when using the restricted
set of metrics available with each anonymization technique.
Further, we reduced the complex ROC curves to a single utility value by computing 
the area under the curve (AUC)~\cite{Bradley1997}. To obtain the 
AUC from the empirical ROC curves, we fitted a piecewise cubic Hermite interpolating 
polynomial to the data, and approximated the area under the curve numerically.
In the next section, we describe and discuss the results obtained with the
methodology described above.

\section{Measurement Results}
\label{sec:results}

We commence this section with an illustrative example for the \emph{loss of resolution} effect, i.e., we examine how the restriction of available resolutions through anonymization impacts the detectability of anomalies. In Fig.~\ref{fig:lossOfResolution}, we plot the count of unique source addresses in all incoming TCP traffic at different resolutions of IP addresses (32), /24 networks (24), /20 networks (20), and /16 networks (16). 
In the curve for the highest resolution value, we see a large peak corresponding to 1.2 million additional IP addresses launching a denial of service attack. Interestingly, the peak completely disappears at a resolution of 24 and lower. Observing the curve for /24 networks more closely, we find a very small, non-significant peak of $\approx$ 5'000 additional networks in the same interval. From this observation, we conclude that the attack sources remain in a few /24 networks. As seen 
in the figure, this example anomaly disappears at lower resolutions, and thus, it will be hard or impossible to detect in data anonymized with truncation. 
In the following, we systematically assess the overall utility of the different anonymization techniques with the help of the Kalman filter detector applied to the whole three-week long data set.

\begin{figure}[t]
  \centering
  \begin{minipage}[b]{0.47\textwidth}
    \centering
	  \includegraphics[scale=0.48]{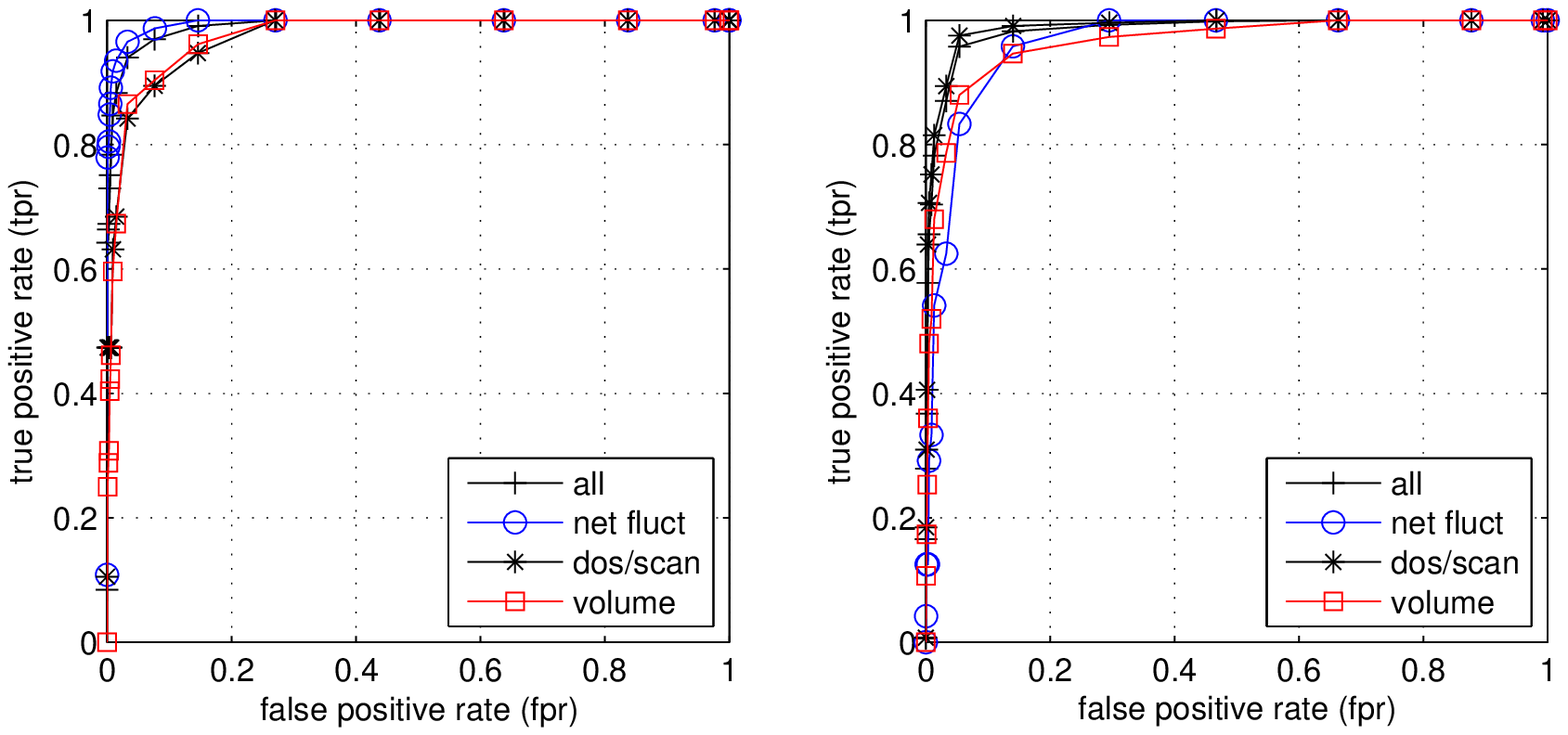}
	  \caption{ROC curves for different types of anomalies in UDP (left) and TCP traffic (right).}
	  \label{fig:rocall}
  \end{minipage}
  \hspace{0.5cm}
  \begin{minipage}[b]{0.47\textwidth}
	  \centering
		 \includegraphics[scale=0.48]{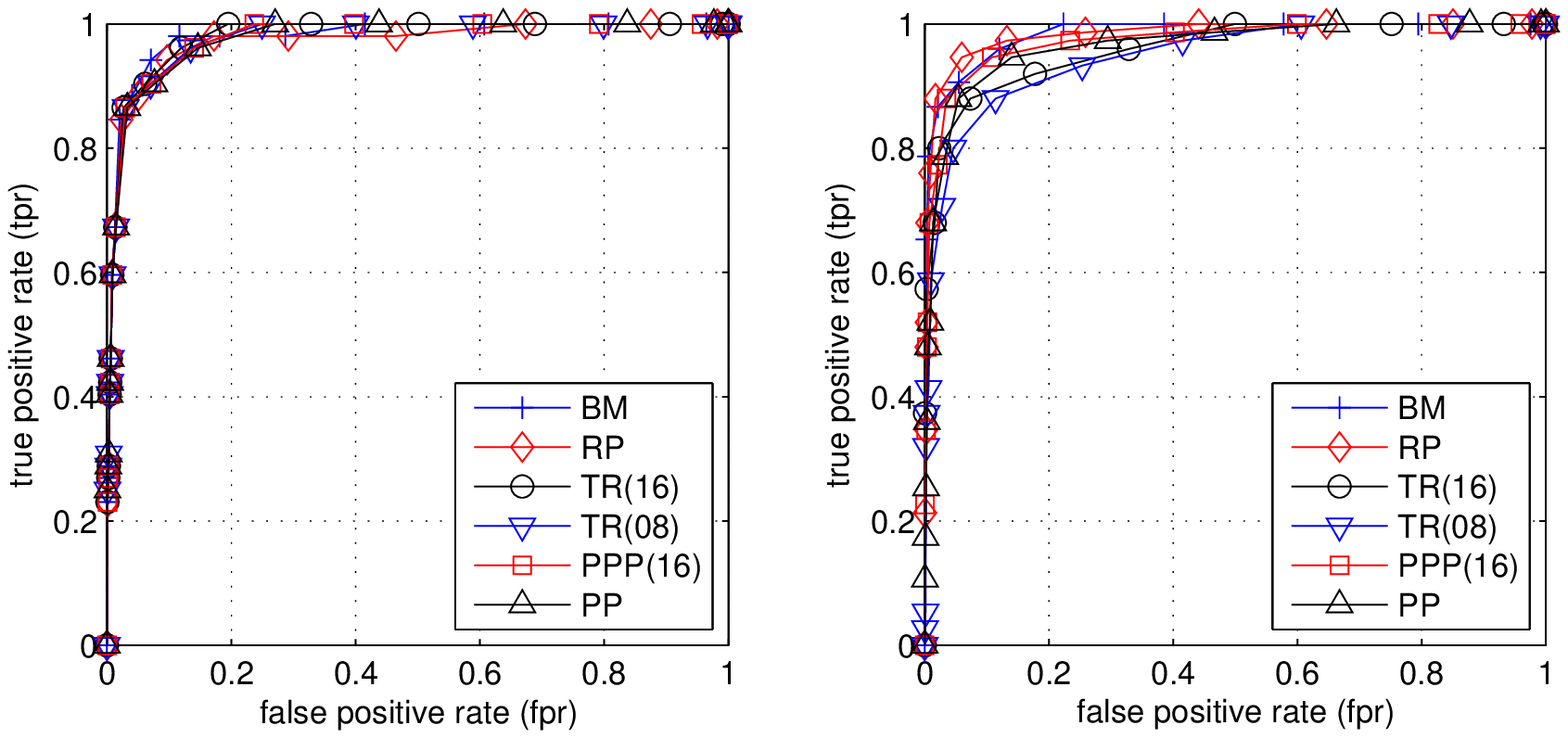}
	  \caption{Volume anomalies in \textit{anonymized} traffic, UDP (left), TCP (right).}
	  \label{fig:rocvol}
	 \end{minipage}
\end{figure}

\begin{figure}[t]
  \centering
  \begin{minipage}[b]{0.47\textwidth}
    \centering
	  \includegraphics[scale=0.48]{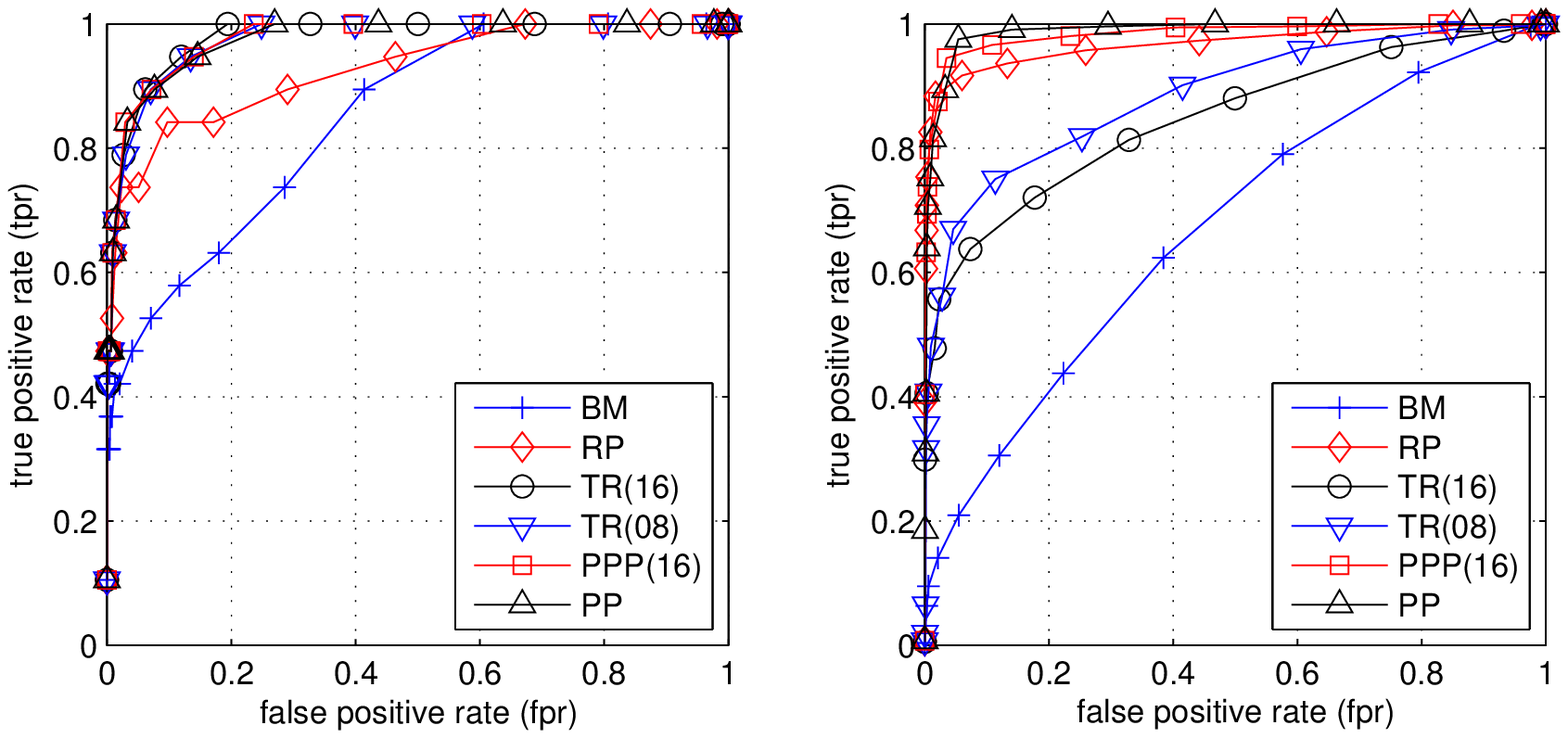}
	  \caption{Scanning and denial of service anomalies in \textit{anonymized} traffic, UDP (left), TCP (right).}
	  \label{fig:rocsd}  
	\end{minipage}
  \hspace{0.5cm}
  \begin{minipage}[b]{0.47\textwidth}
	  \centering
	  \includegraphics[scale=0.48]{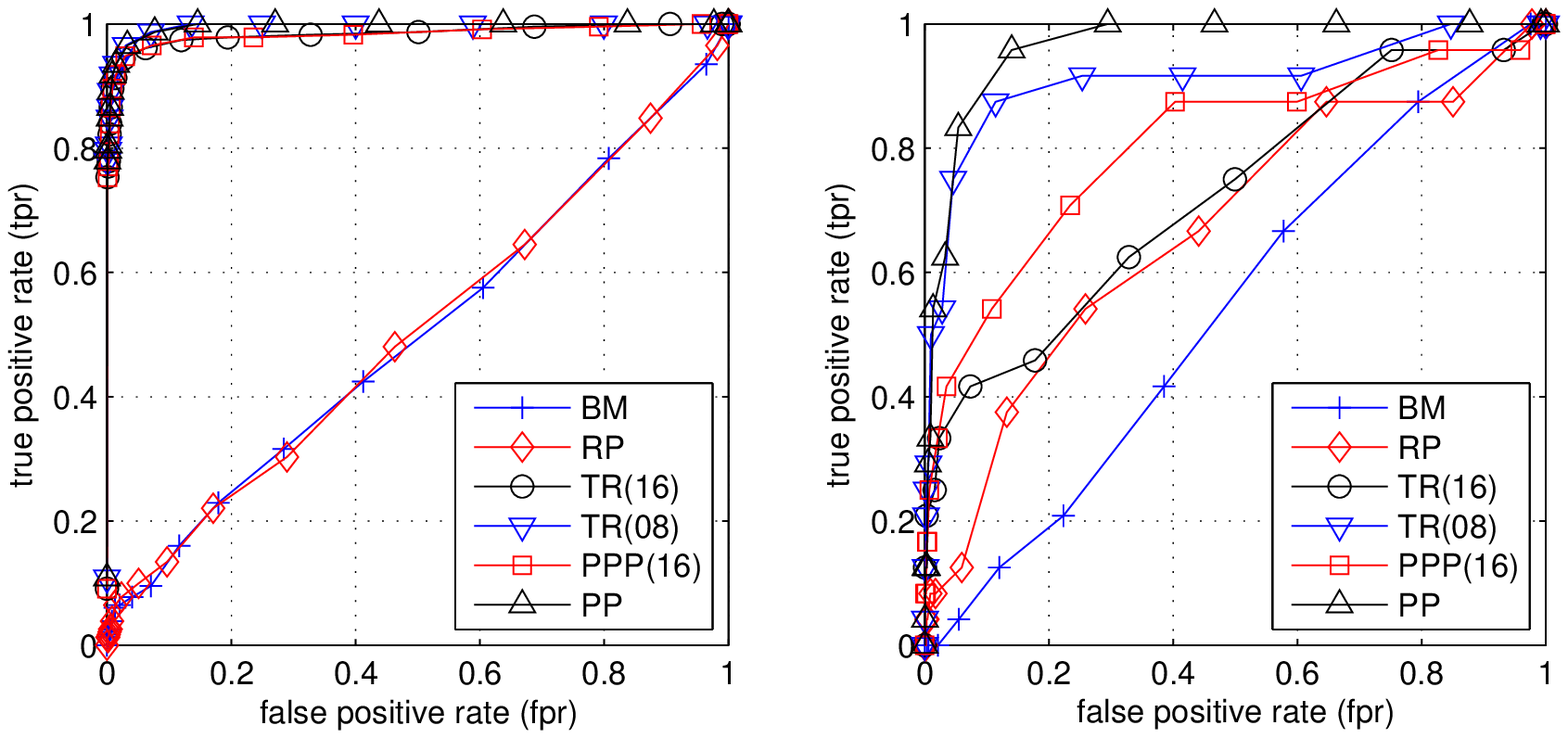}
	  \caption{Network fluctuations in \textit{anonymized} traffic, UDP (left), TCP (right).}
	  \label{fig:rocrou}
	 \end{minipage}
\end{figure}

\subsection{ROC Curves for Anonymized Data}
\label{sec:roc}

To assess the utility of the different anonymization schemes, we study the impact of anonymization separately for each type of anomaly. This is necessary since each type of anomaly is exposed at characteristic traffic granularities, and thus, is 
impacted differently by the restriction of the granularity design space through anonymization.
To give an example: Alpha flows are mainly visible in byte and packet counts computed at the lowest resolution, while scans are primarily visible in feature-based metrics at higher resolutions.
Hence, studying all anomalies together will thus not lead to any conclusive results. As a result, we distinguish the following anomalies for our evaluation:

\emph{Volume Anomalies}, such as outage events and alpha flows, are mainly exposed
by volume-based metrics. Since volume-based metrics at the lowest resolution are available with all 
anonymization schemes (see Table \ref{tab:metrics}), we expect that anonymization does not have a large impact on the detection 
of volume anomalies. Indeed, the measurement 
results presented in Fig.~\ref{fig:rocvol} clearly confirm this expectation. Anonymization 
does not alter the utility of data when one is solely interested in detection of volume anomalies. When examining 
the plot for TCP traffic more closely,
we observe that blackmarking and random permutation perform slightly better than the other schemes. We conclude that using fewer metrics might even be beneficial as it results in fewer false positives.

\emph{Scanning and denial of service anomalies} are both mainly visible in feature-based metrics. Measurement 
results for this class of anomalies are presented in Fig.~\ref{fig:rocsd}. The curves for UDP and TCP confirm that
blackmarking performs worst, whereas prefix-preserving permutation (which does not restrict the granularity design space at all) 
has the best performance. To give an example: at a false positive rate of 0.02, the detection rate for blackmarking is 
reduced by 50\% for UDP, and even more for TCP traffic. Surprisingly, the ranking for truncation and random permutation
is not the same for UDP and TCP traffic. For TCP traffic, random permutation outperforms truncation, while for UDP traffic
the opposite applies. We think this is due to structural differences between, normal and anomalous, UDP and TCP traffic. 
We verified on the data that TCP scans and DoS attacks are mainly visible at the resolution of individual IP addresses, which are preserved by random permutation but not truncation. On the contrary, UDP scans and DoS attacks are visible at high and low resolutions, but metrics at lower resolutions have fewer false positives. Consequently, truncation outperforms random permutation for detecting UDP scans and DoS attacks.

\begin{figure}[t]
	  \centering
	  \includegraphics[scale=0.58,clip]{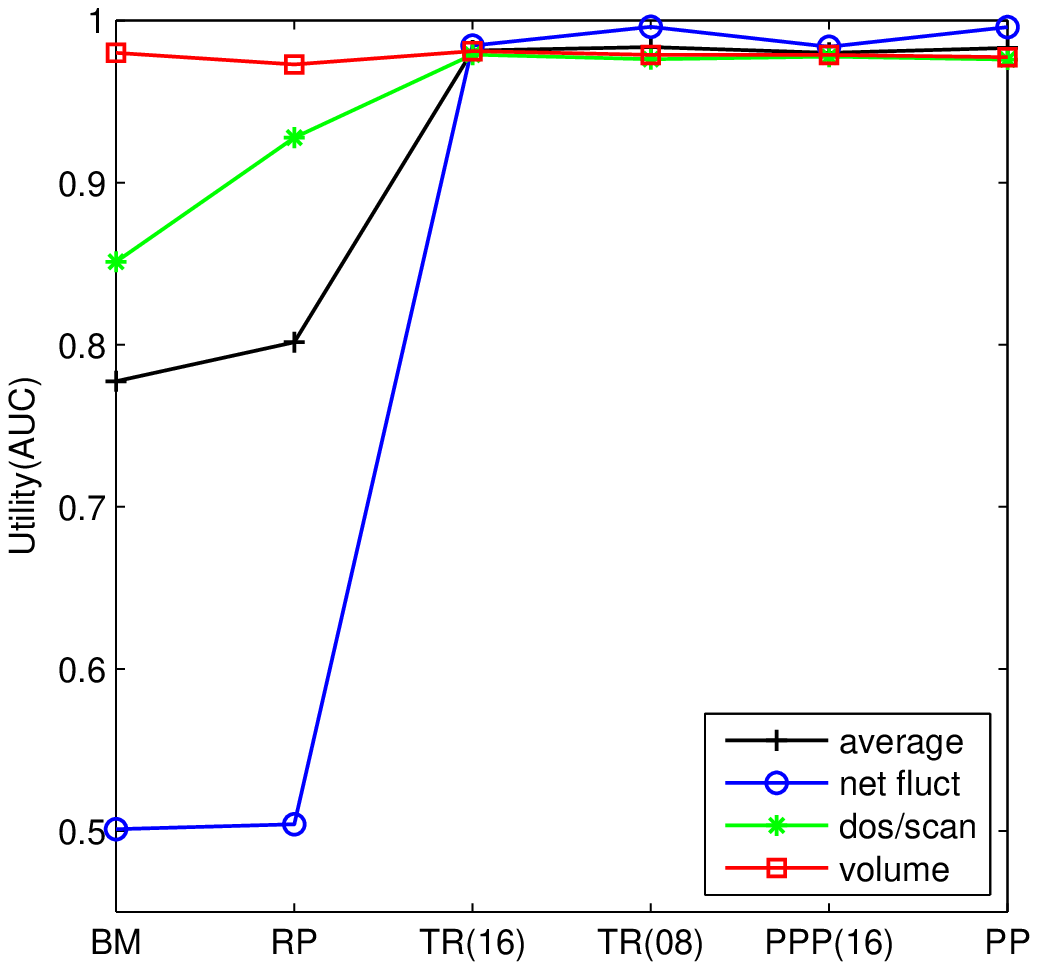}
	  \hspace{0.5cm}
	  \includegraphics[scale=0.58,clip]{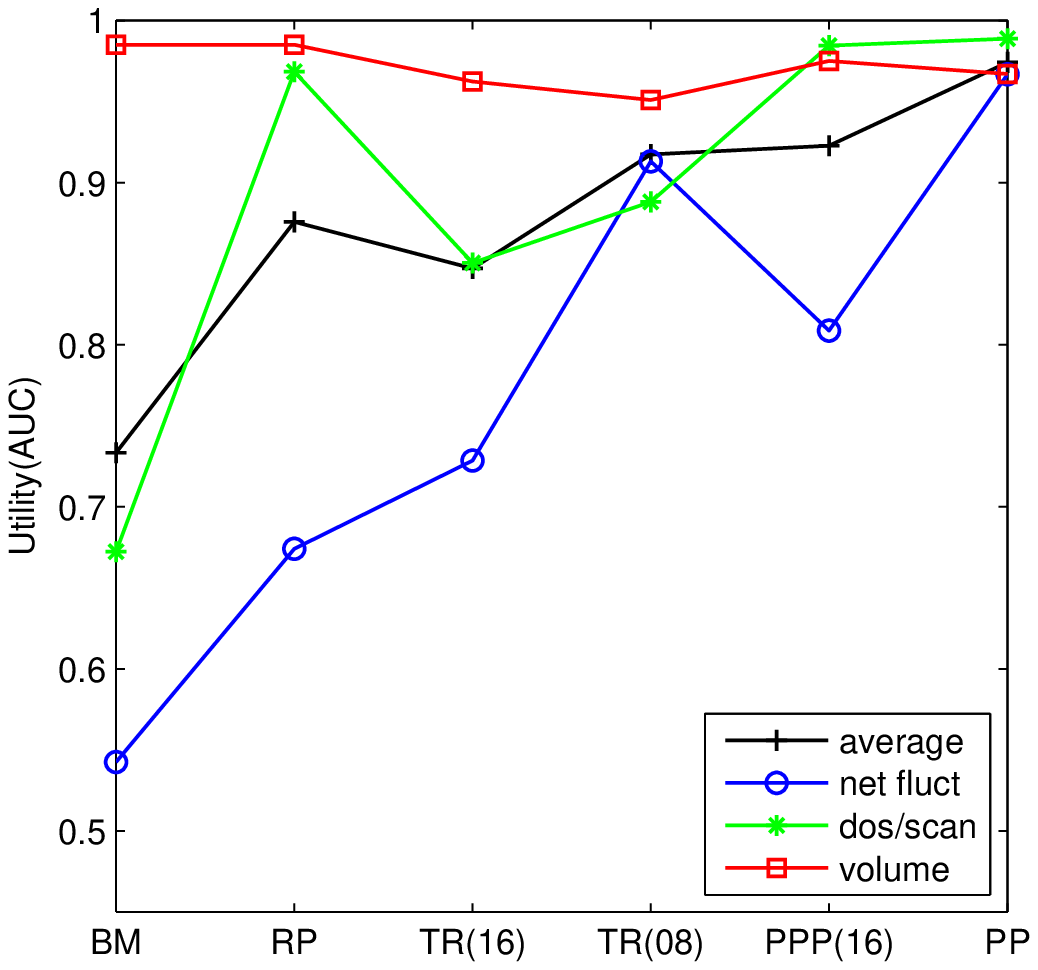}
	  \caption{Utility for large-scale anomaly detection for different anonymization techniques. Left plot is for UDP traffic, right plot is for TCP traffic.}
    \label{fig:auc}

\end{figure}

\emph{Network fluctuations} are mainly visible in feature-based metrics at lower resolutions. The ROC curves for UDP and TCP traffic
presented in Fig.~\ref{fig:rocrou} show that detection of network fluctuations is almost impossible when either blackmarking 
or random permutation is used.
Truncation of 8 bits, on the other hand, does not result in a severe performance degradation, neither for TCP nor UDP traffic. 
UDP and TCP differentiate with respect to the performance of 16-bit truncation and PPP. This might, however, be a particularity of our data set:
Most of the 239 network fluctuation anomalies in the UDP traffic are mainly visible at a resolution of 16 bits, whereas TCP
network fluctuations are rather visible at a resolution of 8 bits.

\subsection{Utility of Anonymized Traces for Anomaly Detection}

We will now summarize the detailed results from Section~\ref{sec:roc} using the area under the curve (AUC)
as a measure for the utility of an anonymized data set. An AUC value of 1 means that the detector achieves perfect
accuracy, whereas a detector with an AUC of 0.5 is not useful.
In Fig.~\ref{fig:auc}, we plot for each anonymization technique (x-axis) the AUC value (y-axis). The left plot
summarizes the results for UDP traffic and the right plot is for TCP traffic.
We show one curve for each of the three anomaly classes (volume, scan/DoS, and network fluctuations), as well as the average over 
all classes. Note that this average utility corresponds to a data set that contains the same ratio of volume, scan/DoS, 
and network fluctuation anomalies.

Fig.~\ref{fig:auc} clearly shows that the detectability of volume anomalies is not impacted by anonymization.
Furthermore, it confirms our intuition that prefix-preserving anonymization has the best utility with respect to 
backbone anomaly detection in general, and that blackmarking has the worst utility for all classes except volume anomalies.
The utility of random permutation, truncation, and partial prefix-preserving permutation largely depends on the type of anomaly 
in question. Partial prefix-preserving permutation performs almost as well as prefix-preserving permutation; it has a lower 
utility only for network fluctuation anomalies in TCP traffic. Random permutation has a high utility for the detection of large-scale scans 
and denial of service attacks, but a low utility for detecting network fluctuations in UDP as well as TCP traffic. 
Truncation performs very well for UDP traffic in our measurements, but has a lower utility for detecting scans, denial of service attacks,
and network fluctuation anomalies in TCP traffic. Moreover, the utility for these anomalies decreases when more bits are truncated.

Naturally, the overall utility of an anonymization technique applied to a given 
data set depends on the mix of anomalies within the trace.
For our data set, the overall utility for detecting TCP anomalies is clearly dominated by the 542 scanning intervals. Likewise, the overall
utility for UDP traffic is to a large extent dominated by the 239 intervals with network fluctuations. Hence, for this anomaly mix, PP and PPP offer 
the highest utility, truncation lies in the middle, and random permutation and blackmarking result in the lowest utility.
We summarize the results for our data set as follows: 
\begin{itemize}
\item Blackmarking decreases the utility for detecting anomalies in UDP and TCP traffic dramatically.
\item Random permutation performs very bad with the detection of anomalies in UDP traffic, while preserving the utility for TCP traffic.
\item Truncation of 8 or 16 bit decreases the utility for detecting anomalies in TCP traffic by roughly 10 percent, while performing well for UDP traffic.
\item (Partial) prefix-preserving permutation has no significant negative impact on the utility for detecting anomalies in UDP and TCP traffic.
\end{itemize}

To derive more general conclusions about the utility of different anonymization schemes it would be helpful 
to study the anomaly mix in available flow traces. If this mix converges, at least for traces recorded in the same 
time period, a general conclusion about the utility 
can be derived directly from our results. Moreover, we expect that the results for volume, scan/DoS, and network fluctuation anomalies hold also for other traces. Verifying this assumption requires that we apply our methodology to further un-anonymized data sets from different networks. Unfortunately, such traces are not currently available in abundance.

\section{Implicit Traffic Aggregation}
\label{sec:example}

In Section~\ref{sec:results}, we investigated how anomaly detection results are falsified 
when valuable resolutions are not available for anomaly detection. 
Another aspect of anonymization that is worth studying is the restriction 
along the subset size dimension, causing an implicit aggregation of traffic. 

Let us illustrate this with an example.
Consider the case where traffic from a single host (i.e., subset size 32)
is to be investigated in presence of 4 bits truncation. The best one can do 
under these circumstances is investigation of the /28 network that contains the host, since the individual
host can no longer be distinguished from other hosts in the same network. As a consequence,
the analyzed traffic is a mixture of traffic from the target host and traffic from other hosts in neighboring
subnets. In accordance with loss of resolution, we refer to this effect as the \emph{loss of focus effect}. 

The impact of the loss of focus effect is, of course, highly
dependent on the distribution of traffic in the studied network. 
\emph{It is extremely difficult to predict the implications for a particular case.} 
In the worst case, even truncation of a single bit can be fatal. That is, traffic characteristics of the
target host could get lost completely in another host's traffic.
In the best case, where the truncated subnet is dedicated to a single host, no traffic is aggregated. In that case, the
loss of focus effect is simply reduced to a loss of resolution effect.

We can, however, estimate the average severity of truncation-induced traffic aggregation
by analyzing the count of additional, non-belonging, flows for 170 individual hosts. 
In particular, we count the flows of 170 webservers belonging to a single /16 network in our un-anonymized traces. 
Then, we apply truncation to the traces and count the number of flows to the subnets containing the webservers.

When truncating a single bit, more than 50\% of the observed webservers experience no additional traffic.
Only around 10\% of the webservers have a resulting traffic increase of
100\% or more. However, if more bits are truncated the situation
gets worse. The ratio of unaffected servers drops to 20\% for 2 bits, 5\%
for 4 bits, and even 0\% for 8 bits. Similarly, the ratio of servers that
experience at least a doubling of traffic goes up to 25\% for 2 bits,
55\% for 4 bits and 89\% for 8 bits.

We conclude that accurate detection of small-scale anomalies\footnote{With small-scale anomalies we denote anomalies affecting only a single host or a small subnet.} is very difficult, if not impossible, when the desired subset size is not supported by the restricted granularity design space. The probability that host characteristics are lost in aggregated traffic is simply too high.
Only anomalies of sufficient scale have a chance to be visible in aggregated traffic at larger subset sizes. In addition, false positives are introduced by the aggregation with traffic from other hosts.
\section{Conclusion}
\label{sec:conclusion}

In this paper, we have answered the question of how anonymization techniques impact statistical anomaly detection. 
We introduced the detection granularity design space as an important tool to illustrate this impact. 
We have shown how the design space, spanned by subset size and resolution, 
is reduced by the most common IP address anonymization techniques. Finally, we analyzed the utility of anonymized traces 
for the detection of large-scale anomalies, as well as the impact on traffic characteristics of individual hosts with 
the aid of backbone traffic traces gathered over a three-weeks period.

In general, our results indicate that the restriction of the granularity design space through anonymization hinders anomaly detection. With respect to the individual techniques, we have found that prefix-preserving permutation offers the best utility, and blackmarking performs the worst. Moreover, we have shown that the performance of random permutation, partial-prefix preserving permutation, and truncation strongly depends on the type of anomaly that is studied, as well as the underlying transport protocol. 
Our results indicate that the detection of volume anomalies, such as outages or alpha flows, is not impacted 
by anonymization at all. The utility for detecting scans and denial of service attacks degrades
when truncation is applied. Detection of network fluctuations, on the other hand, is impacted principally
by blackmarking and random permutation.

Thus, if one is interested in a particular type of anomaly, anonymization could be tuned in a way such that the results are less impaired. 
In addition, we have shown that the anonymization-induced loss of focus, i.e., when the desired subset size is not available, in most cases completely distorts the traffic characteristics of individual hosts. 

While we provided some interesting insights on the impact of anonymization techniques on anomaly detection, we encourage further research with un-anonymized traffic traces to challenge or confirm our results.

\section*{Acknowledgments}
We are grateful to Elisa Boschi from Hitachi Europe for the numerous valuable discussions and to SWITCH for providing the traffic traces used in this study.

\section*{REFERENCES}
\footnotesize
\bibliographystyle{acm}

\end{document}